\documentclass{jpp}
\usepackage{graphicx,color} 
\usepackage[utf8]{inputenc}
\usepackage[T1]{fontenc}
\usepackage{amsmath}

\usepackage{lipsum}
\usepackage{upgreek}
\usepackage{gensymb}
\usepackage{hyperref}
\DeclareMathOperator\sinc{sinc}

\shorttitle{Towards Laser Ion Acceleration With Holed Targets}
\shortauthor{P. Hadjisolomou, S. V. Bulanov and G. Korn}

\title{Towards Laser Ion Acceleration With Holed Targets}

\author{Prokopis Hadjisolomou\aff{1}
  \corresp{\email{Prokopis.Hadjisolomou@eli-beams.eu}},
  S. V. Bulanov\aff{1,2}
 \and G. Korn\aff{1}}

\affiliation{
\aff{1}Institute of Physics of the ASCR, ELI-Beamlines project, Na Slovance 2, 18221 Prague, Czech Republic
\aff{2}National Institutes for Quantum and Radiological Science and Technology (QST), Kansai Photon Science Institute, 8-1-7 Umemidai, Kizugawa, Kyoto 619-0215, Japan
}

\begin{document}

\maketitle

\begin{abstract}
\par Although the interaction of a flat-foil with currently available laser intensities is now considered a routine process, during the last decade emphasis is given to targets with complex geometries aiming on increasing the ion energy. This work presents a target geometry where two symmetric side-holes and a central-hole are drilled into the foil. A study of the various side-holes and central-hole length combinations is performed with 2-dimensional particle-in-cell simulations for polyethylene targets and a laser intensity of $5.2 \kern0.1em {\times} \kern0.1em 10^{21} \kern0.2em \mathrm{W \kern0.1em cm^{-2}}$. The holed-targets show a remarkable increase of the conversion efficiency, which corresponds to a different target configuration for electrons, protons and carbon ions. Furthermore, diffraction of the laser pulse leads to a directional high energy electron beam, with a temperature of ${\sim} \kern0.1em 40 \kern0.2em \mathrm{MeV}$ or seven times higher than in the case of a flat-foil. The higher conversion efficiency consequently leads to a significant enhancement of the maximum proton energy from holed-targets. 
\end{abstract}


\section{Introduction} \label{Introduction}

\par During the last decades the Chirped Pulse Amplification method \citep{Strickland1985, Mourou2006} enabled laser systems to undergo a rapid power increase. Once the laser intensity exceeds ${\sim} \kern0.1em 10^{18} \kern0.2em \mathrm{W \kern0.1em cm^{-2}}$ then an isotropic proton emission is observed from the laser interaction with a flat-foil \citep{Maksimchuk2000}, with proton energies exceeding $1 \kern0.2em \mathrm{MeV}$. Recently, two groups approached the $100 \kern0.2em \mathrm{MeV}$ proton energy threshold \citep{Higginson2018, Kim2016} by using a ${\sim} \kern0.1em 5 \kern0.1em {\times} \kern0.1em 10^{20} \kern0.2em \mathrm{W \kern0.1em cm^{-2}}$ intensity laser but on the cost of using nanometre-thickness targets which hardly survive the interaction with the finite contrast prepulse of \mbox{multi-PW} laser systems available in the near future \citep{Rus2011, Zamfir2012, Bashinov2014, Zou2015}.

\par In contrast with conventionally accelerated proton beams, laser-driven proton beams contain a high number of particles generated in a time comparable to the laser pulse duration, although the divergence of the beam is of ${\sim} \kern0.1em 30 \degree$ full-angle \citep{Daido2012, Macchi2013}. However, the reduced cost of laser-driven proton beams make them extremely attractive on several existing and/or proposed applications, including the fast ignition of thermonuclear fusion targets \citep{Roth2001, Atzeni2002}, hadron therapy \citep{Bulanov2002b, Bulanov2004b, Murakami2008, Bulanov2014, Kurup2019}, isotope production \citep{Santala2001}, material sciences \citep{Booty1996} and proton imaging \citep{Borghesi2002, Borghesi2003}. Specifically for the cancer therapy, ${\sim} \kern0.1em 250 \kern0.2em \mathrm{MeV}$ protons are required to treat a patient, setting a socially accepted minimum proton energy boundary.

\par A laser beam that mimics the planned High-Repetition-Rate Advanced Petawatt Laser System (HAPLS) in the Extreme Light Infrastructure (ELI) Beamlines \citep{Rus2011}, Czech Republic is assumed for studying the processes discussed in this work. It is estimated that HAPLS will deliver $30 \kern0.2em \mathrm{J}$ energy within $27 \kern0.2em \mathrm{fs}$ pulse duration at Full Width Half Maximum (FWHM), resulting in a peak power of $1 \kern0.2em \mathrm{PW}$, while the central wavelength, $\lambda$, of the laser radiation equals to $815 \kern0.2em \mathrm{nm}$. The focal spot diameter is assumed $4.75 \kern0.2em \mathrm{\upmu m}$ FWHM which corresponds to a peak intensity of $5.2 \kern0.1em {\times} \kern0.1em 10^{21} \kern0.2em \mathrm{W \kern0.1em cm^{-2}}$. The resulting amplitude of the maximum electric field is $1.98 \kern0.1em {\times} \kern0.1em 10^{14} \kern0.2em \mathrm{V \kern0.1em m^{-1}}$ (corresponding to a dimensionless amplitude, $\alpha_0$, of ${\sim} \kern0.1em 50$) and its spatial extent is $6.72 \kern0.2em \mathrm{\upmu m}$ FWHM, equals to ${\sim} \kern0.1em 8 \lambda$.


\subsection{Acceleration Mechanisms} \label{Acceleration_Mechanisms}

\par The most well-known laser driven ion acceleration mechanism is the Target Normal Sheath Acceleration (TNSA) \citep{Wilks2001, Daido2012, Passoni2010, Macchi2013} which dominates at intensities of $10^{20} \kern0.2em \mathrm{W \kern0.1em cm^{-2}}$ and requires the incidence of a finite contrast laser pulse on a thick (in the micrometre range) flat-foil, where the electron number density of the target is higher than the critical electron number density, $n_{cr}$. Due to the laser-foil interaction, a hot electron population is created with a fraction of it permanently escaping the target. As a consequence of the electron ejection, a longitudinal electric field is established on the target rear surface along with an expanding plasma cloud. Since protons are present on the target as contaminants, they are favourably accelerated due to their high charge to mass ratio, resulting in a spectrum characterised by a double temperature and a sharp cut-off energy.

\par Once the laser intensity reaches $10^{21} \kern0.2em \mathrm{W \kern0.1em cm^{-2}}$ the Radiation Pressure Acceleration (RPA) mechanism \citep{Esirkepov2004, Klimo2008, Robinson2008, Bulanov2016} becomes important, where its experimental indications have already been observed \citep{Higginson2018, Kim2016}. The RPA mechanism is based on the concept of the relativistic mirrors, where the radiation pressure of the laser pulse pushes the foil (in the interaction region) as a whole. The energy of the pulse is transferred to the foil due to reflection by the co-moving foil mirror. The RPA mechanism requires a circularly polarised laser to be effective, while for a $1 \kern0.2em \mathrm{PW}$ laser then nanometre-thick foils are required \citep{Higginson2018, Kim2016}. One drawback of the RPA mechanism is that it requires high laser contrast for the foil not to be destroyed by the prepulse prior the arrival of the main pulse. In the present work only linear polarisation is used and the RPA mechanism is not involved.

\par If the laser is able to extract almost all electrons from a foil region, then the Coulomb Explosion (CE) acceleration mechanism \citep{Morita2008, Bulanov2008, Ogura2012} occurs. Several requirements must be met for CE to be effective, such as a high laser contrast and an optimum target thickness, depending the pulse parameters and target density. In order of the CE mechanism to occur, electrons must be ejected within a duration defined by the electron ejection time and the ion response time. Approximately, only electrons within the focal spot contribute to CE. The sudden electron ejection leaves the target with a positively charged ion core. The strong repulsive forces between ions lead to an explosion of the ion core, where the ion spectrum is characterised by a non-thermal distribution. The maximum ion energy is proportional to the maximum electrostatic potential value within the ion core prior the explosion.

\par If the target has an electron number density near the critical electron number density for relativistically intense electromagnetic waves, $n_{cr}^{rel}$, then the incident pulse can channel through the target setting the basis for the Magnetic Vortex Acceleration mechanism (MVA) \citep{Bulanov2007, Yogo2008, Bulanov2010, Park2019}. Once the pulse exits the target then a quasi-static magnetic field is established, related with the vortex trajectories of electrons in that region. The resulting magnetic field sustains a longitudinal electric field, which is responsible for ion acceleration.


\subsection{Structured Targets and Holed-Target Geometry} \label{Geometry}

\par Not long after the first MeV laser generated protons were demonstrated by laser irradiation of flat-foils, simulations proposed that a curved foil could collimate the proton beam \citep{Bulanov2000, Sentoku2000, Wilks2001} and an experimental demonstration came soon after \citep{Patel2003}, setting the new path of laser irradiated microstructured targets. A few notable target proposals include a foil doped with a proton-rich region \citep{Bulanov2002b, Esirkepov2002, Schwoerer2006}, a hollow microsphere \citep{Burza2011}, foils with solid microspheres attached on their front surface \citep{Klimo2011, Margarone2012}, foils attached to hollow cones \citep{Bartal2012,Qiao2013}, foils with a series of microslits \citep{Nagashima2014, Kawata2014}, atomic clusters \citep{Ditmire1997}, microwires developed on the target front surface \citep{Jiang2016, Rocca2017, Bargsten2017} and laser irradiated microtubes/microchannel-plates \citep{Ji2016, Ji2019, Snyder2019}. Fabrication of microstructures is now considered a routine process and features of ${\sim} \kern0.1em 100 \kern0.2em \mathrm{nm}$ can be created by direct laser writing \citep{Fischer2013}.

\par
\begin{figure}
  \centering
  \includegraphics[width=0.80\linewidth]{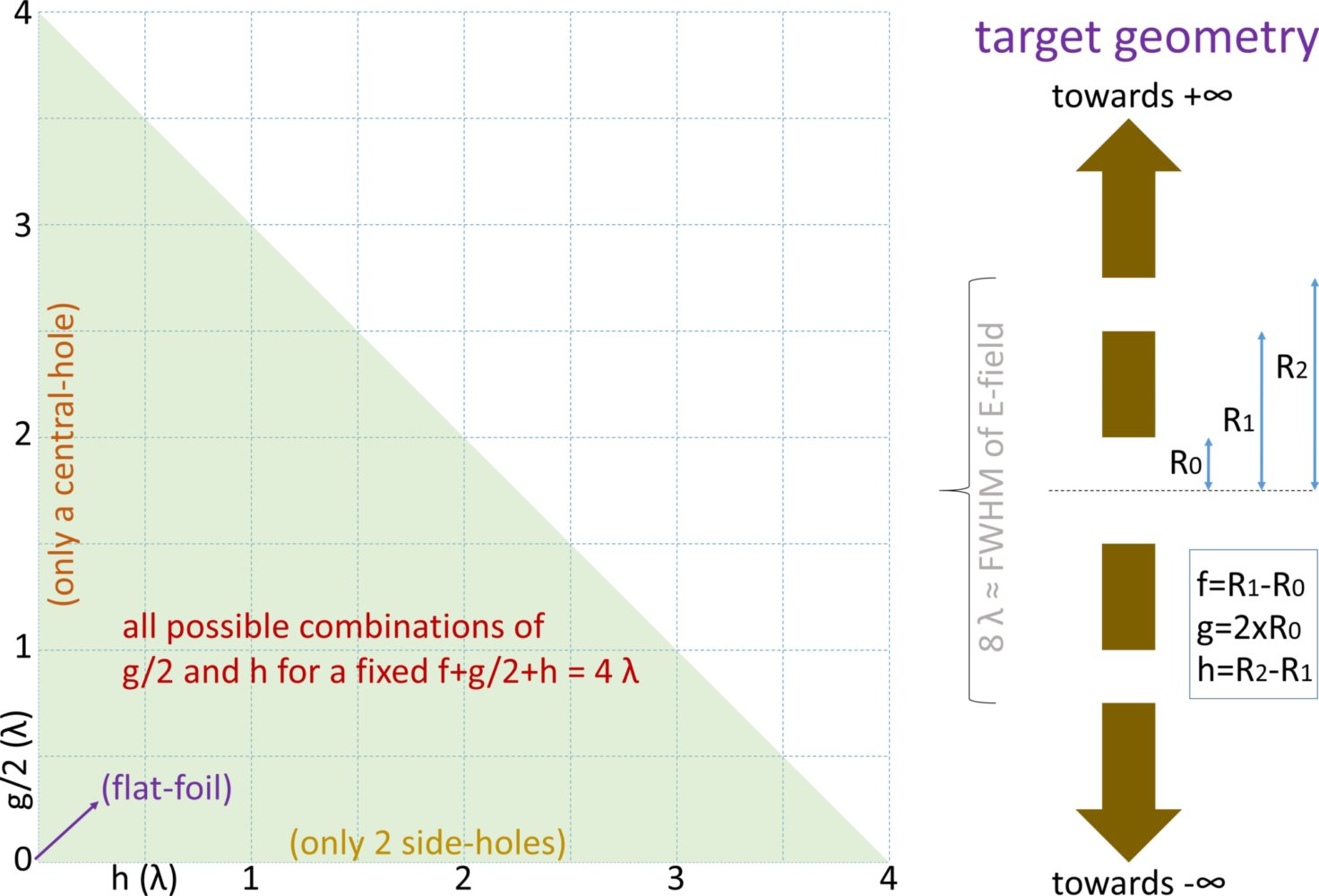}
  \caption{
  (\textit{left}) The green orthogonal triangular area contains the combinations of the central-hole radius and the side-holes extent, where the length sum of the central-hole, the two side-holes and the two inner-segments equals $8 \kern0.1em \lambda$. The base and the height of the triangle correspond to HT2SH and HT1CH respectively. The intersection of the two cathetus corresponds to a flat-foil and the hypotenuse is a sub-case of the holed-target geometry where the two inner-segments are removed. (\textit{right}) Schematic representation of a sampled holed-target used in the simulations. The target consists of a central-hole, defined as $g$, two side-holes, each represented by $h$, two inner-segments, each noted as $f$, and two outer-segments which extent to infinity.
  }
\label{fig:1}
\end{figure}

 Here we propose an innovative holed-target geometry where a central-hole is drilled into a $1 \kern0.2em \mathrm{\upmu m}$ thick polyethylene \citep{Bamberger1900} foil, centred on the laser focal spot. In addition, two side-holes are drilled on the two-dimensional (2D) foil geometry, symmetrically to the central-hole, as shown in the right side of Fig. \ref{fig:1}. The central-hole radius, side-hole inner edge and side-hole outer edge correspond to radii of $R_0$, $R_1$ and $R_2$ respectively, as labelled in Fig. \ref{fig:1}. The foil segment corresponding to a length of $R1-R_0$ is called inner-segment (defined as $f$) and the segment extending from $R_2$ to $\infty$ is called outer-segment. The extent of the central-hole radius (defined as $g/2$), the side-hole extent (defined as $h$) and the inner-segment extent equal a length of $4 \lambda$, meaning that the electric field amplitude at the interaction point and at $R_2$ have a ratio of 2. The targets considered in the present work cover all combinations of central-hole radius and side-hole extent. From these combinations, since the length sum of the target components (ignoring the outer-segment) is $8 \lambda$ the effect of the inner-segment length can also be extracted. Although the target holes are adjusted to a focal spot with dimensions usually used in laser-proton acceleration experiments, a larger spot \citep{Steinke2019} allows more flexibility on the dimension of the holes.

\par A graphical representation of all possible holed-target geometries if the condition $f + g/2 + h = 8 \lambda$ is met is shown in the left side of Fig. \ref{fig:1}, where only the green triangle corresponds to allowed combinations of the central-hole and the side-holes extent. Notably, the general holed-target geometry corresponds to two significantly simplified geometries, one on the base and one on the height of the triangle. If one considers only the combinations on the base of the triangle then the target transforms to a centred mass-limited (ML) target plus an outer-segment, named as holed-target with 2 side-holes (HT2SH); if the combinations on the height of the triangle are considered then the target corresponds to a flat-foil with a hole in the centre, named as holed-target with 1 central-hole (HT1CH). The intersection of the base and height of the triangle correspond to a flat-foil. All cases along the hypotenuse of the triangle correspond to the same case of a holed-target (where the inner-segments are removed) regardless of the value of either $g$ or $h$ as long as their sum is $4 \lambda$. The only exception is presented in Sec. \ref{ML} for a HT2SH, where the central foil radius is kept at $0.5 \lambda$ and the side-holes' extent is gradually increased.

\section{Simulations} \label{Setup}

\par The results of this work are obtained through the 2D version of EPOCH \citep{Arber2015} Particle In Cell (PIC) code. The EPOCH version $4.17.19$ is compiled with the Higuera-Cary \citep{Higuera2017, Ripperda2018} instead of the default Boris solver \citep{Ripperda2018, Boris1970}. Protons with a velocity exceeding $0.5 \kern0.2em c $ (where $c$ is the speed of light) and electrons approaching $c$ are expected. Therefore, the use of the Higuera-Cary solver is necessary in to simultaneously preserve volume and the ${ {\bf{E}} \kern0.1em {\times} \kern0.1em {\bf{B}} }$ velocity \citep{Higuera2017}. In the present work, although a spectra comparison of the two solvers exhibits minor differences for protons, the electron spectra are significantly different.

\par A combination of three overlapping particle species is used, namely electrons, protons (H) and carbon ions (C), consisting a polyethylene foil with a mass density of $0.93 \kern0.2em \mathrm{g \kern0.1em cm^{-3}}$ and a chemical formula of $\mathrm{(C_2 H_4)}_n$, where the subscript $n$ is an integer and denotes a macroscopic chain. Electrons, protons and C are assigned to a mass of $1, 1836.15$ and $12 \kern0.1em {\times} \kern0.1em 1836.15$ times the electron rest mass, a charge of $-1, 1,$ and $6$ times the elementary charge and a number density of $3.2 \kern0.1em {\times} \kern0.1em 10^{23} \kern0.2em \mathrm{cm^{-3}}$, $8 \kern0.1em {\times} \kern0.1em 10^{22} \kern0.2em \mathrm{cm^{-3}}$ and $4 \kern0.1em {\times} \kern0.1em 10^{22} \kern0.2em \mathrm{cm^{-3}}$ respectively. All holed-targets are $1 \kern0.2em \mathrm{\upmu m}$ thick, with the location and extent of the three holes varying in each simulation.

\par The simulations use ``open'' boundaries in all directions apart the negative (left) \mbox{x-boundary}, which is set to ``simple-laser'' allowing the launch of a laser pulse from it. The ``open'' boundaries allow both fields and particles reaching the box boundary to freely exit the simulation, although the simulation box is large enough to ensure that neither particles nor fields escaped within the simulation time. The laser pulse is launched along the target normal (along $\bf{\hat{x}}$), following the focusing equations of a Gaussian beam \citep{Siegman1986}, focused  at the $(0,0)$. The laser is P-polarised with a wavelength of $815 \kern0.2em \mathrm{nm}$. The electric field on the focus is described by a Gaussian spatial profile of $\sqrt{2} \kern0.1em {\times} \kern0.1em 4.75 \kern0.2em \mathrm{\upmu m}$ FWHM (equals ${\sim} \kern0.1em 8 \lambda$) and a temporal profile of $27 \kern0.2em \mathrm{fs}$ FWHM (equals ${\sim} \kern0.1em 10 \lambda$), where the temporal profile is delayed by two standard deviations in order of the rising pulse profile to be accurately generated. All simulations use a peak intensity of $5.2 \kern0.1em {\times} \kern0.1em 10^{21} \kern0.2em \mathrm{W cm^{-2}}$.

\par The simulations are performed on the ECLIPSE cluster, using for each simulation 36 nodes with 16 processors per node, where these processors are distributed to a square arrangement of $24 \kern0.1em {\times} \kern0.1em 24$. To avoid numerical heating by the open boundaries, the dimensions of the simulation box in each direction are set to $122.88 \kern0.2em \mathrm{\upmu m} \kern0.1em {\times} \kern0.1em 122.88 \kern0.2em \mathrm{\upmu m}$. The length of each cell in each direction is $10 \kern0.2em \mathrm{nm}$, which approximately equals the skin depth of polyethylene by ignoring the correction due to the relativistically intense electromagnetic waves, while by considering the relativistic correction then the skin depth becomes more than five times larger than the cell size ($56 \kern0.2em \mathrm{nm}$). Therefore, each dimension corresponds to $3 \kern0.1em {\times} \kern0.1em 2^{12}$ cells and $9 \kern0.1em {\times} \kern0.1em 2^{25}$ macroparticles are assigned to each specie. By considering that macroparticles are assigned only to cells with non-zero density, initially each cell contains ${\sim} \kern0.1em 40$ macroparticles, with the exact number depending on the holes of each holed-target. By considering that the laser pulse needs ${\sim} \kern0.1em 200 \kern0.2em \mathrm{fs}$ to reach the focal spot in the centre of the box, during the initial $160 \kern0.2em \mathrm{fs}$ evolution of particles is disabled to approximately half the computational time. Each simulation runs for $400 \kern0.2em \mathrm{fs}$ with $21199$ steps (time-step multiplier factor of 0.8), with a time-step equal ${\sim} \kern0.1em 1.77$ times the cell size over $c$.


\section{Results and Discussion}

\par This section provides a detailed description of the results of the general holed-target geometry simulations. Initially, Sec. \ref{DF} explains how the initial Gaussian pulse profile is rearranged in space after diffracted and/or reflected by a holed-target. Following, Sec. \ref{EEEPD} emphasises on electron ejection and propagation following the laser-target interaction, along with the resulting charge separation. In addition, Sec. \ref{EEEPD} quantifies the amount of laser energy transferred to the target particles. In Sec. \ref{EEPD} acceleration of protons and carbon ions is studied for the general holed-target configuration. Emphasis on the electron and proton spectra extracted from two limit cases of  HT2SH and HT1CH is given in Sec, \ref{Spectra}. The results close with Sec. \ref{ML}, where a transition from the holed-target geometry to a ML target is made.


\subsection{Diffracted Fields} \label{DF}

\begin{figure}
  \centering
  \includegraphics[width=1.00\linewidth]{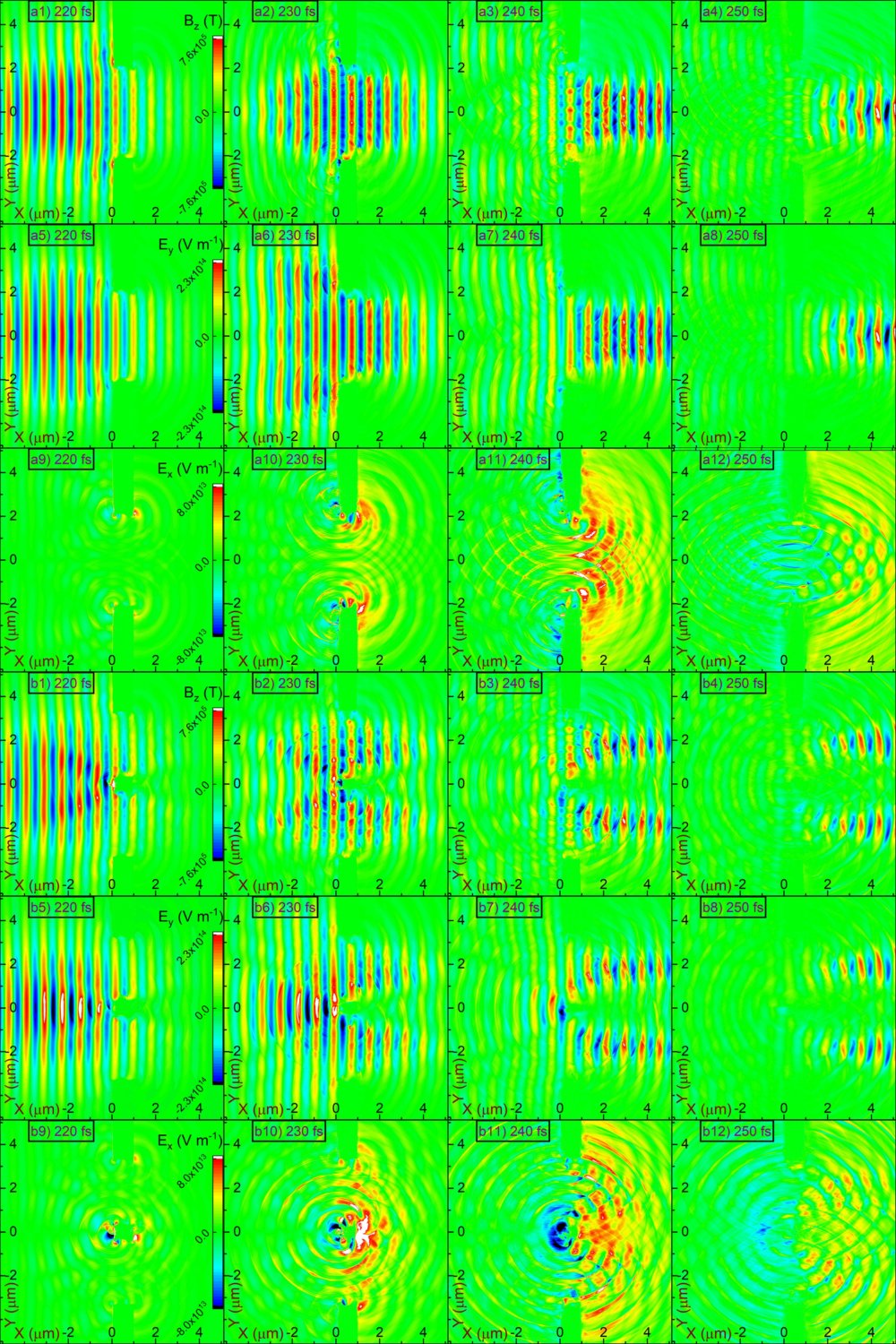}
  \caption{Evolution of the magnetic ($B_z$) and electric ($E_y$ and $E_x$) field components corresponding to a
  (\textit{a1-a12}) HT1CH with a central-hole of $5 \kern0.1em \lambda$ and (\textit{b1-b12}) HT2SH with two side-holes of $3.5 \kern0.1em \lambda$.
  }
\label{fig:6_Exy}
\end{figure}

\par Every laser interaction with a general target geometry results in the same laser energy rearrangement, where a portion of the laser energy is transferred to the target particles (Sec. \ref{EEEPD}), a small portion is captured by the target as a surface field \citep{Quinn2009, Tokita2015}, a third portion is reflected and a fourth portion is transmitted towards the laser propagation direction. In general, the holed-target geometry consists of an infinite flat-foil on which three gaps are created. Therefore, it can be described by the theory of diffraction and interference of a planar wave by multiple slits \citep{Born1999}.

\par However, although a theoretical model can predict the evolution of the laser field after the interaction with the polyethylene foil, several assumptions make a precise analytical prediction hardly possible. Although analytical solutions of the diffracted intensity do exist for an infinite planar wave incident on multiple slits, in our case the pulse has a spatial Gaussian envelope. Furthermore, during the laser-target interaction the target geometry is gradually altered, meaning that the final field configuration is not a result of a single diffraction and interference behaviour. Therefore, the complex field configuration is accurately obtained only through simulations.

\par The resulting field evolution is shown in Fig. \ref{fig:6_Exy}, where the upper half corresponds to a HT1CH with a central-hole of $5 \lambda$ and the lower half to a HT2SH with two side-holes of $3.5 \lambda$. In both cases, due to the large extent of the holes (large fraction of the pulse spatial profile), most of the laser pulse is transmitted through the target forming an approximately predictable diffraction pattern. The extent of the diffraction effect can be better visualised by considering the longitudinal component of the electric field, $E_x$, which although initially absent, it appears once the pulse passes through the holed-target. Notably, a strong $E_x$ field is created in the area between the central-hole of the HT1CH. The diffracted field intensity, $I$, is approximated by the diffraction of a plane wave from a single slit,
\begin{equation}
 I = I_0 \left[ \sinc \left( \frac{\upi \kern0.1em \alpha}{\lambda} \sin(\theta) \right) \right]^2
\end{equation}
Here, $I_0$ is the intensity of the assumed plane wave, $\alpha$ is the slit opening and $\theta$ is the angle at which the diffracted intensity is estimated. In agreement with the low divergence in Fig. \ref{fig:6_Exy}, for $5 \lambda$ the intensity at FWHM is associated with an angle of ${\sim} \kern0.1em 5^o$. Additionally, an even stronger $E_x$ appears on the rear surface of the inner-segment of the HT2SH. Notably, $E_x$ in both cases has a duration approximately equal the laser pulse duration.

\par Since the laser spatial extent is larger than the holes extent, a portion of the pulse is reflected by the front target surface. The reflection is not identical for all holed-target configurations, resulting in various complex interference patterns. The final picture is further complicated since the interference can be either constructive or destructive as time evolves. In the case of a HT1CH and a large hole to wavelength ratio, interference is not important since practically only the edges of the pulse are reflected. However, for a HT2SH, the central part of the pulse faces a solid inner-segment. Therefore, if the inner-segment is large enough, the pulse is reflected in a manner similar to reflection from a flat-foil. The destructive interference of the magnetic field and constructive interference of the electric field can be seen in Fig. \ref{fig:6_Exy}(b2) and Fig. \ref{fig:6_Exy}(b6) respectively, where a dimmer region of the magnetic field is accompanied by a brighter region of the electric field.


\subsection{Energy Transfer, Electron Evolution and Charge Separation} \label{EEEPD}

\begin{figure}
  \centering
  \includegraphics[width=1.00\linewidth]{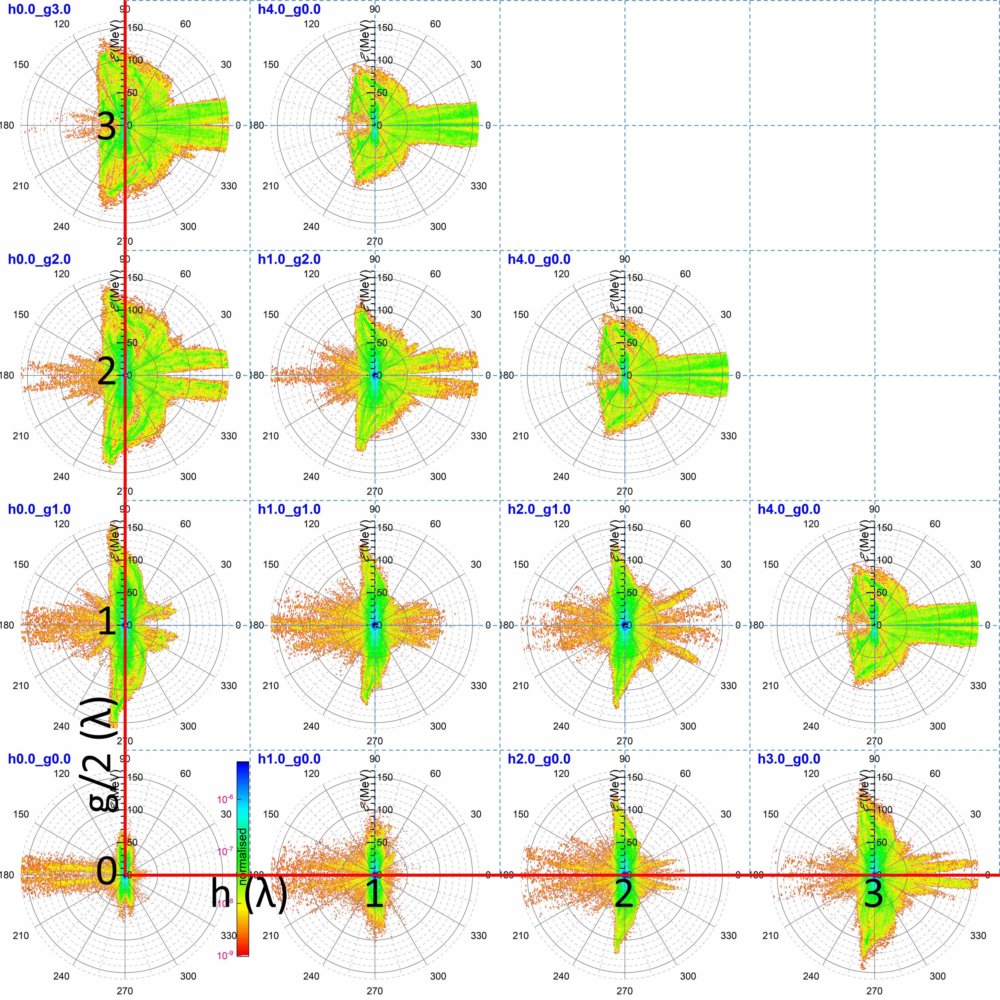}
  \caption{Electron energy polar diagrams for various combinations of the central-hole radius and the side-holes extent. The red lines form a Cartesian system where the holed-target characteristic dimensions ($h$ and $g/2$) are given by the centre of each contour.
  }
\label{fig:2e}
\end{figure}

\begin{figure}
  \centering
  \includegraphics[width=1.00\linewidth]{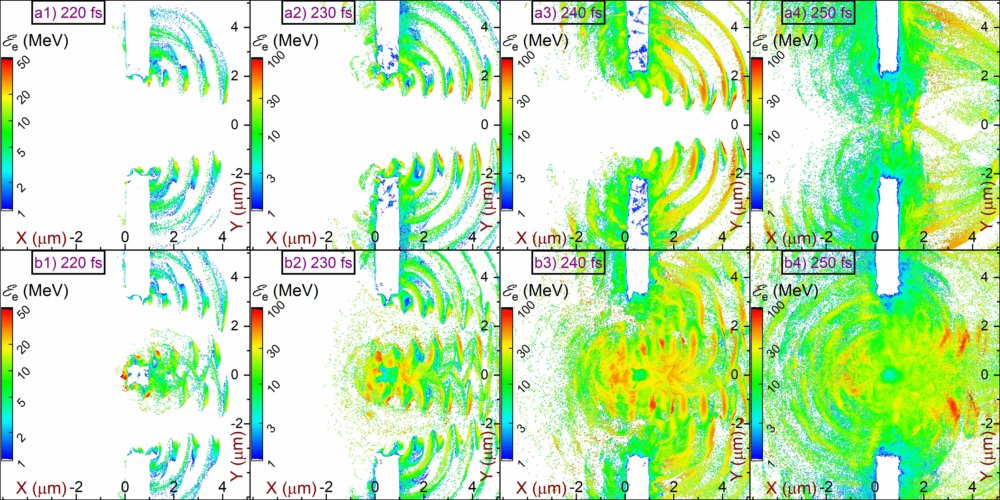}
  \caption{Evolution of the electron mean energy per cell corresponding to a (\textit{a1-a4}) HT1CH with a central-hole of $5 \kern0.1em \lambda$ and (\textit{b1-b4}) HT2SH with two side-holes of $3.5 \kern0.1em \lambda$.
  }
\label{fig:4a}
\end{figure}

\begin{figure}
  \centering
  \includegraphics[width=1.00\linewidth]{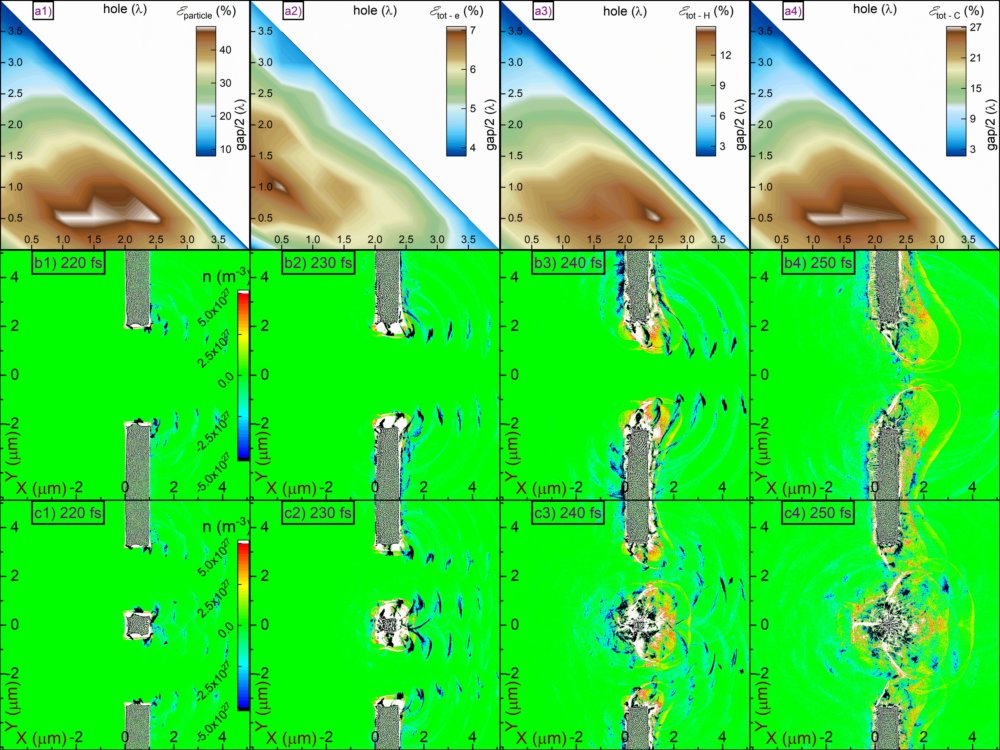}
  \caption{The percentage of laser energy transferred to (\textit{a1}) all particles, (\textit{a2}) electrons, (\textit{a3}) protons and (\textit{a4}) carbon ions by the end of the simulation. Evolution of the charge density per cell for a (\textit{b1-b4}) HT1CH with a central-hole of $5 \kern0.1em \lambda$ and (\textit{c1-c4}) HT2SH with two side-holes of $3.5 \kern0.1em \lambda$.
  }
\label{fig:5a}
\end{figure}

\par When the rising front of the laser pulse (assumed zero for distances greater than two standard deviations from the pulse peak) reaches the holed-target, it is reflected at the locations where inner-segments and outer-segments exist. However, due to the initially low field amplitude of the pulse rising front (approximately one order of magnitude less than the peak), the field penetrates the target at a depth approximately given by the skin depth of polyethylene (it is equal to ${\sim} \kern0.1em 10 \kern0.2em \mathrm{\upmu m}$) creating a hot electron population. At a time of ${\sim} \kern0.1em 23 \kern0.2em \mathrm{fs}$ the electric field on target gets its peak value, penetrating at ${\sim} \kern0.1em 56 \kern0.2em \mathrm{\upmu m}$ within the target, further heating electrons.

\par The initial hot electron distribution eventually leads to several distinct populations. The largest population corresponds to a hot electron cloud refluxing the target, where this behaviour is known as electron ``stochastic heating'' \citep{Bulanov2015, Yogo2015, Yogo2017}. In our case, these electrons are visualised as the small blue dot in Fig. \ref{fig:2e}, at the centre of each subfigure. Another electron fraction moves along the foil in both directions, following a portion of the laser field trapped by the foil and propagating outwards. These electrons have been observed experimentally in cases where the laser pulse had a large incidence angle on a flat-foil and a strong electron signal was observed as originating from the foil edges \citep{Li2006, Mao2012, Wang2013}. A similar effect was observed by wires directly interacting with a laser pulse \citep{Tokita2011, Nakajima2013, Tokita2015}.

\par A third electron fraction escapes the target as seen by the highest energy electrons in Fig. \ref{fig:4a}. Notably, the spatial distribution of the mean electron kinetic energy partially overlaps with the spatial distribution of $E_x$, as seen by comparing Fig. \ref{fig:2e} with Fig. \ref{fig:6_Exy}. This effect is mentioned as ``electron capture'' by the laser field \citep{Wang1998, Zhy1998, Wang2001, Braenzel2017}, where similar electron patterns are observed in Refs. \citet{Bulanov2002, Bulanov2010a}. Therefore, those captured electrons are continuously accelerated for as long as they are kept in phase with the pulse \citep{Wang2001, Bulanov2002}.

\par For HT1SH the electron mean energy contours (see Fig. \ref{fig:4a}(a)) do not completely correspond to the field spatial distributions (see Fig. \ref{fig:6_Exy}(a)). The partial overlap of the two patterns is due to the lack of a inner-segment, since the produced electrons originate only from the edges of the pulse. On the other hand, for HT2SH, a plethora of electrons is ejected from the inner-segment. That behaviour is also related to the different extents of the hot electron cloud. If no inner-segment exists in the path of the pulse then less electrons recirculate the interaction region, as indicated by the reduced size of the central blue dot at the first column and outer diagonal in Fig. \ref{fig:2e}, where it is also realised by comparing Fig. \ref{fig:4a}(a4) with Fig. \ref{fig:4a}(b4).

\par The distinction of the several electron populations in combination with the holed-target geometries explain the form of all subfigures grouped in Fig. \ref{fig:2e}. In the case of a flat-foil (see the intersection of the cathetus in Fig. \ref{fig:2e}), most of the laser pulse is reflected, capturing a portion of hot electrons driving them into hundreds of MeV energies. Furthermore, the portion of the pulse captured by the foil drives electrons along the foil surface at an energy of ${\sim} \kern0.1em 50 \kern0.2em \mathrm{MeV}$.

\par The first column of Fig. \ref{fig:2e} corresponds to HT1CH, where the hole extent is increasing towards the hypotenuse. As the hole extent increases, smaller amplitude fields are reflected, decreasing the energy and number of the target front electrons. On the other hand, the percentage of the diffracted pulse increases, preserving the initial shape of the pulse due to weaker diffraction by the increased hole. As a result, the target rear electrons are accelerated to hundreds of MeV energies. Notably, this behaviour has an optimum hole extent, since for an extremely large hole a drastic lowering of the pulse amplitude allows no electron ejection.

\par The last row of Fig. \ref{fig:2e} corresponds to HT2SH, where the side-holes extent is increasing towards the hypotenuse. Practically, by increasing the side-holes extent the inner-segment extent is decreased and, a smaller portion of the pulse is reflected by the inner-segment. For small inner-segments electrons are preferably accelerated towards the target rear, but not as strongly as in the case of HT1CH. This is because the inner-segment (located in the middle of the incident pulse) reflects the highest field amplitude but it also diffracts the remaining pulse significantly attenuating it.

\par Ejection of hot electrons from the target leads to a charge imbalance, as seen in Fig. \ref{fig:5a}. The blue regions indicate an excess of negative charge, due to the capture of electrons by the laser field. However, the white (saturated) regions indicate an excess of protons and carbon ions, which eventually leads to CE. The positive core is stronger for a HT2SH with two side-holes of ${\sim} \kern0.1em 3.5 \kern0.1em \lambda$ since at that case most electrons are ejected from the inner-segment at ${\sim} \kern0.1em 230 \kern0.2em \mathrm{fs}$, as can be realised by the large positively charged region in Fig. \ref{fig:5a}(c2). The conditions under which CE dominates are given in Egs. 1 and 2 of Ref. \cite{Bulanov2016}, which can be rearranged as
\begin{equation}
 l < \frac{2 \kern0.1em \varepsilon_0 \kern0.1em E}{e \kern0.1em n_e} .
\end{equation}
Here, $l$ is the foil thickness, $\varepsilon_0$ is the permittivity of free space. $E$ the electric field, $e$ the elementary charge and $n_e$ the electron number density. For a HT2SH with a central-segment of $\lambda$, due to the $7 \lambda$ total gap in the FWHM of the electric field, the target thickness should be 8 times larger since $l$ in the equation above is volumetrically associated with the number of hot electrons. Therefore, for that specific case one should expect complete electron ejection at a central-segment thickness of ${\sim} \kern0.1em 0.55 \kern0.2em \mathrm{\upmu m}$. However, due to the complex diffraction patterns the pulse is able to penetrate the central-segment and extract electrons from both its front and rear surfaces, further relaxing the minimum target thickness required.

\par The percentage of the laser energy transferred to particles is also directly connected with the geometry of the holed-target, as seen in Fig. \ref{fig:5a}(a1). Although for a flat-foil only ${\sim} \kern0.1em 25 \kern0.2em \%$ of the energy is transferred to particles, this value approximately doubles for a holed-target with a central-hole of $0.5 \kern0.1em \lambda$ and side-holes of $1-2.5 \kern0.1em \lambda$. The drop of the conversion efficiency for larger holes is because then less particles interact with the laser pulse. Furthermore, the conversion efficiency of each particle specie is maximised for different holed-targets. For electrons (see Fig. \ref{fig:5a}(a2)), a clear peak is formed for a central-hole of $1 \kern0.1em \lambda$ and side-holes of $0.5 \kern0.1em \lambda$, while for protons (see Fig. \ref{fig:5a}(a3)) this value shifts to $0.5 \kern0.1em \lambda$ for central-hole and $2.5 \kern0.1em \lambda$ for side-holes. As for carbon ions (see Fig. \ref{fig:5a}(a4)), since they define most energy absorbed by particles, their optimum absorption corresponds to the same hole values as those of the total system, but with a slight optimisation for side-holes of $1 \kern0.1em \lambda$.


\subsection{Ion Energy Evolution and Polar Distribution} \label{EEPD}

\begin{figure}
  \centering
  \includegraphics[width=1.00\linewidth]{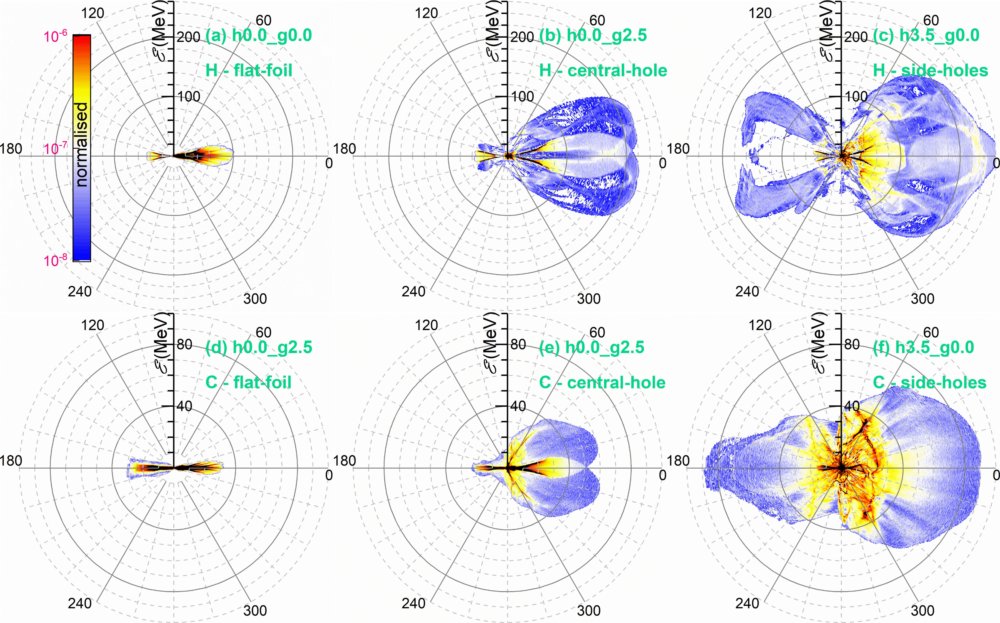}
  \caption{Proton energy polar diagrams corresponding to a (\textit{a}) flat-foil, (\textit{b}) HT1CH with a central-hole of $5 \kern0.1em \lambda$ and (\textit{c}) HT2SH with two side-holes of $3.5 \kern0.1em \lambda$. Carbon ion energy polar diagrams corresponding to a (\textit{d}) flat-foil, (\textit{e}) HT1CH with a central-hole of $5 \kern0.1em \lambda$ and (\textit{f}) a HT2SH with two side-holes of $3.5 \kern0.1em \lambda$.
  }
\label{fig:3f}
\end{figure}

\begin{figure}
  \centering
  \includegraphics[width=1.00\linewidth]{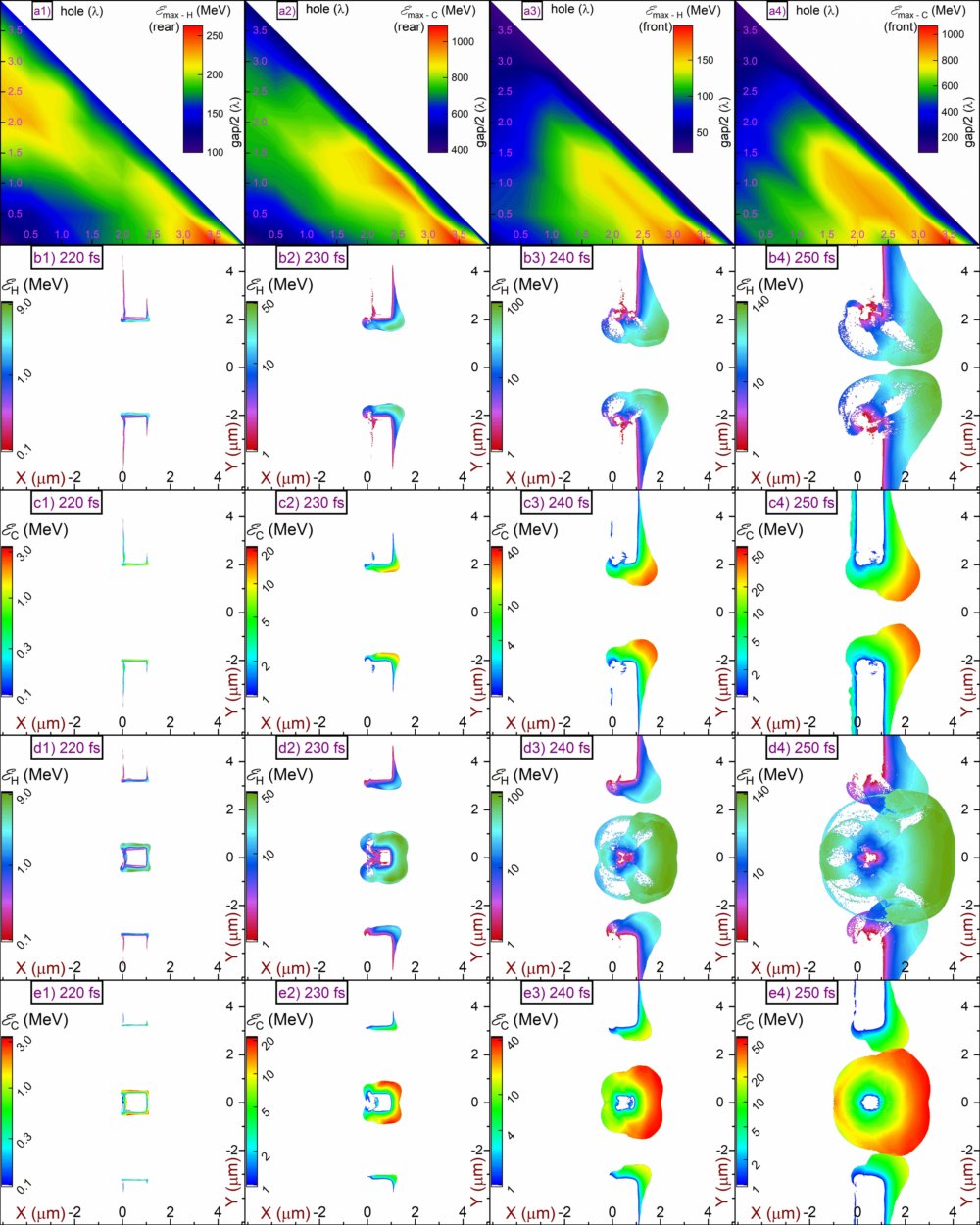}
  \caption{The maximum energy of holed-target (\textit{a1}) rear protons, (\textit{a2}) front protons, (\textit{a3}) rear carbon ions and (\textit{a4}) front carbon ions, by the end of the simulation. Evolution of the mean kinetic energy per cell for (\textit{b1-b4}) protons and (\textit{c1-c4}) carbon ions, for a HT1CH with a central-hole of $5 \kern0.1em \lambda$. Evolution of the mean kinetic energy per cell for (\textit{d1-d4}) protons and (\textit{e1-e4}) carbon ions, for a HT2SH with two side-holes of $3.5 \kern0.1em \lambda$.
  }
\label{fig:5b}
\end{figure}

\par A direct consequence of electron ejection from the target is acceleration of protons and carbon ions and a brief report on the main laser driven ion acceleration mechanisms is made in Sec. \ref{Acceleration_Mechanisms}. As described in Sec. \ref{Geometry}, the general holed-target geometry corresponds to two limiting cases (HT1CH and HT2SH), where a flat-foil is the common geometry of these limits. Therefore, the holed-target geometry does not follow a single acceleration mechanism, and for different target parameters different acceleration mechanisms dominate.

\par In the unique case of the flat-foil, the proton acceleration is due to the TNSA mechanism. In our case the target is a $1 \kern0.2em \mathrm{\upmu m}$ polyethylene foil and the pulse is P-polarised, excluding any contribution of the RPA mechanism at the present intensity. Furthermore, the solid foil density does not allow the MVA mechanism to develop. Therefore, the simulations for a flat-foil give only ${\sim} \kern0.1em 100 \kern0.2em \mathrm{MeV}$ protons and ${\sim} \kern0.1em 30 \kern0.2em \mathrm{MeV}$ per nucleon carbon ions, as shown in Fig. \ref{fig:3f}. The divergence of the beam is ${\sim} \kern0.1em 30 \degree$ full angle, as expected from the TNSA mechanism \citep{Macchi2013}.

\par Let us first consider the geometrically simplest limiting case, of a HT1CH. To understand how protons are accelerated from this configuration, one must consider the proton and carbon ion mean energy evolution, shown in  Fig. \ref{fig:5b}(b) and Fig. \ref{fig:5b}(c) respectively, along with the evolution of the field components, shown in Fig. \ref{fig:6_Exy}(a). Initially, the edges of the Gaussian pulse interact with the outer-segments resulting in electron ejection. Therefore, a sheath field is created on the outer-segments (practically they are two independent foils) weakly accelerating protons by a TNSA-like mechanism; since the acceleration is normal on the foil surfaces (both target rear and inner), it appears as protons originating from the target corners, as shown in Fig. \ref{fig:5b}(b2, c2). Furthermore, electron ejection leads to accumulation of a localised net positive charge (weak CE), as seen in Fig. \ref{fig:4a}(b2). As a result, protons reach the target rear region, as seen in Fig. \ref{fig:5b}(b3, c3). There, diffraction of the laser pulse leads to development of a longitudinal electric field (see \mbox{Fig. \ref{fig:6_Exy}(a7-a12)}) which then accelerates the protons, as seen in Fig. \ref{fig:5b}(b4, c4). Since the diffracted longitudinal electric field exists only in the target rear region, protons in the target front region are only weakly accelerated due to the initial weak CE acceleration. The simulations show that although protons are emitted with a divergence of ${\sim} \kern0.1em 90 \degree$ full angle, their vast majority is confined within ${\sim} \kern0.1em 30 \degree$, as shown in Fig. \ref{fig:3f}(b).

\par The acceleration mechanism governing the second holed-target limiting case, of HT2SH, is the well-examined CE mechanism, where the expected maximum proton energy is given by ${\cal{E}}_p \approx 173 \sqrt{ {\cal{P}} [\mathrm{PW}]} \kern0.2em \mathrm{MeV}$ (${\cal{P}} [\mathrm{PW}]$ is the laser power in $\mathrm{PW}$) \citep{Bulanov2004b, Bulanov2014} and for the present intensity corresponds to ${\sim} \kern0.1em 180 \kern0.2em \mathrm{MeV}$. The strong charge imbalance observed in Fig. \ref{fig:5a}(c3) indicates that the CE regime has been reached. Another indication of the CE mechanism is the similar energies of both the target front and target rear protons, as shown in Fig. \ref{fig:3f}(c); the same applies for carbon ions, seen in Fig. \ref{fig:3f}(f). Again, although protons have an overall divergence of ${\sim} \kern0.1em 90 \degree$, an increased proton number is observed within a ${\sim} \kern0.1em 30 \degree$ full angle. The simulations suggest a maximum proton energy of ${\sim} \kern0.1em 260 \kern0.2em \mathrm{MeV}$, or ${\sim} \kern0.1em 30 \kern0.2em \%$ higher that what predicted in Ref. \citet{Bulanov2014}. However, the holed-targets do not obey a single acceleration mechanism, and additional acceleration due to the diffracted longitudinal electric field component (although not the dominant mechanism) must be acknowledged, as in the HT1CH case. Notably, if the inner-segment is large enough to reflect most of the laser pulse, then electrons are not effectively ejected and the inner-segment does not undergo strong CE. In that case, the situation resembles an intermediate regime of CE and TNSA, where the TNSA dominates for larger inner-segments. Therefore, by increasing the inner-segment extent the maximum proton energy is reduced, approaching the energy obtained from a flat-foil.

\par The results for the many geometrical combinations of the holed-targets are visually represented in Fig. \ref{fig:5b}(a1-a4), where the maximum proton and carbon ion energy is plotted as a function of the side-holes and central-holes dimensions, for both the target front and rear surface. From Fig. \ref{fig:5b}(a1) it is observed that two local maxima exist on the contour of the rear surface proton energy (Fig. \ref{fig:5b}(a1)), corresponding to the transition of the general holed-target geometry to the two limiting cases. The highest energy-maximum corresponds to the HT2SH. Since at that geometry CE dominates, an energy-maximum exist for the same geometric configuration (same contour location) in Fig. \ref{fig:5b}(a3) corresponding to the target front protons, but also in Figs. \ref{fig:5b}(a2) and \ref{fig:5b}(a4) corresponding to carbon ions. The extent of the inner-segment suggests an optimum proton energy value, where all electrons are ejected from the inner-segment leaving an ion core. However, if the inner-segment is further reduced, the ion core becomes smaller resulting in a weaker CE. On the other hand, for the case of a HT1CH, the energy-maximum is observed only in Fig. \ref{fig:5b}(a1) due to the absence of the longitudinal electric field in the target front region; the energy-maximum is weaker in Fig. \ref{fig:5b}(a2) due to the lower charge to mass ratio of carbon ions compared to protons.


\subsection{Electron and Ion Spectra from Holed-Targets} \label{Spectra}

\begin{figure}
  \centering
  \includegraphics[width=1.00\linewidth]{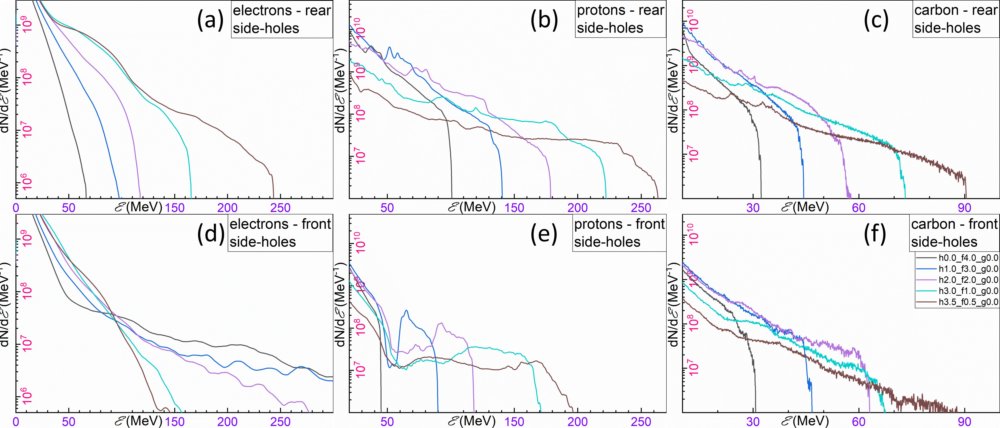}
  \caption{Energy spectra of (\textit{a}) electrons, (\textit{b}) protons and (\textit{c}) carbon ions corresponding to the rear of a HT2SH. Energy spectra of (\textit{d}) electrons, (\textit{e}) protons and (\textit{f}) carbon ions corresponding to the front of a HT2SH. The target geometry is defined by the $g$ and $h$ parameters as labelled in subfigure (f), while the black line corresponds to a flat-foil.
  }
\label{fig:10_horizontal}
\end{figure}

\begin{figure}
  \centering
  \includegraphics[width=1.00\linewidth]{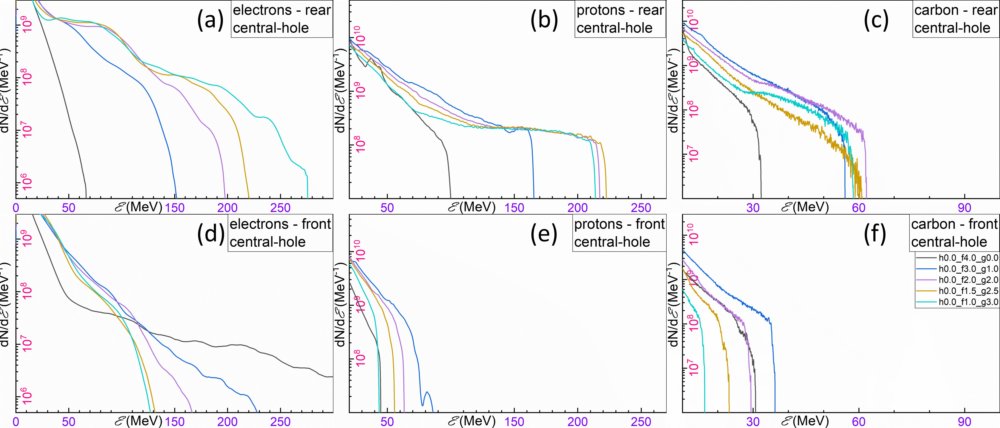}
  \caption{Energy spectra of (\textit{a}) electrons, (\textit{b}) protons and (\textit{c}) carbon ions corresponding to the rear of a HT1CH. Energy spectra of (\textit{d}) electrons, (\textit{e}) protons and (\textit{f}) carbon ions corresponding to the front of a HT1CH. The target geometry is defined by the $g$ and $h$ parameters as labelled in subfigure (f), while the black line corresponds to a flat-foil.
  }
\label{fig:10_vertical}
\end{figure}

\par The qualitative results of Fig. \ref{fig:5b}(a1-a4) are quantified in Figs. \ref{fig:10_horizontal} and \ref{fig:10_vertical}, where spectra for electrons, protons and carbon ions are extracted for both limiting holed-target geometries. The spectra from the 2D simulations have been rescaled to the equivalent 3D spectra \citep{Hadjisolomou2020} assuming a $4.75 \kern0.2em \mathrm{\upmu m}$ emission region diameter.

\par The energy gain of the several hot electron populations is explained in Sec. \ref{EEEPD} and pictured in Fig. \ref{fig:2e}. However, the target rear electron spectra from holed-targets reveal a significantly enhanced flux compared to flat-foils (black line in Figs. \ref{fig:10_horizontal} and \ref{fig:10_vertical}); notably, the opposite behaviour occurs for target front electrons. The flat-foil simulations suggest a hot electron temperature of ${\sim} \kern0.1em 5.5 \kern0.2em \mathrm{MeV}$; for the intensity used, Haines scaling \citep{Haines2009} gives ${\sim} \kern0.1em 4 \kern0.2em \mathrm{MeV}$ and Kluge scaling \citep{Kluge2011} gives ${\sim} \kern0.1em 7 \kern0.2em \mathrm{MeV}$. On the other hand, the holed-target geometry suggests an electron temperature of ${\sim} \kern0.1em 30 \kern0.2em \mathrm{MeV}$ for HT2SH and ${\sim} \kern0.1em 40 \kern0.2em \mathrm{MeV}$ for HT1CH (brown lines in Figs. \ref{fig:10_horizontal}(a) and \ref{fig:10_vertical}(a) respectively).

\par The target rear proton spectra for a HT2SH is shown in Fig. \ref{fig:10_horizontal}(b). Although the cut-off proton energy is also shown in Fig. \ref{fig:5b}(a), the spectra reveal the reduction of the total number of accelerated protons for increased maximum proton energy due to shrink of the inner-segment from which the high energy protons originate from. Despite the proton number reduction, a continuously increasing temperature is observed for increasing maximum proton energy (see Fig. \ref{fig:10_horizontal}(b)). Therefore, the envelope of all cases of maximum proton energy appears with an exponentially-like decaying behaviour. A similar trend is also seen for carbon ions, as shown in Fig. \ref{fig:10_horizontal}(c).

\par On the other hand, the target rear proton spectra for HT1CH exhibit several qualitative differences. The maximum proton energy follows an increasing trend with increasing the central-hole extent, up to ${\sim} \kern0.1em 2 \lambda$. Then, for an increasing central-hole up to ${\sim} \kern0.1em 3 \lambda$ the maximum proton energy remains practically unchanged, with a maximum at ${\sim} \kern0.1em 2.5 \lambda$. Following that maximum region, further increase in the central-hole extent drastically reduces the maximum proton energy. Regardless the target geometry, all spectra are characterised by a double exponential decaying function and a sharp cut-off energy (see Fig. \ref{fig:10_vertical}(b)). Notably, the volumetric reduction of the number of accelerated protons is not obvious, since the accelerated protons originate from the perimeter of the central-hole. For a large central-hole, although a larger perimeter suggests a larger number of protons available to be accelerated, a weaker combination of TNSA and CE drives less protons into the region of the accelerating component of the diffracted electric field.


\subsection{Towards Mass-Limited Targets} \label{ML}

\begin{figure}
  \centering
  \includegraphics[width=0.60\linewidth]{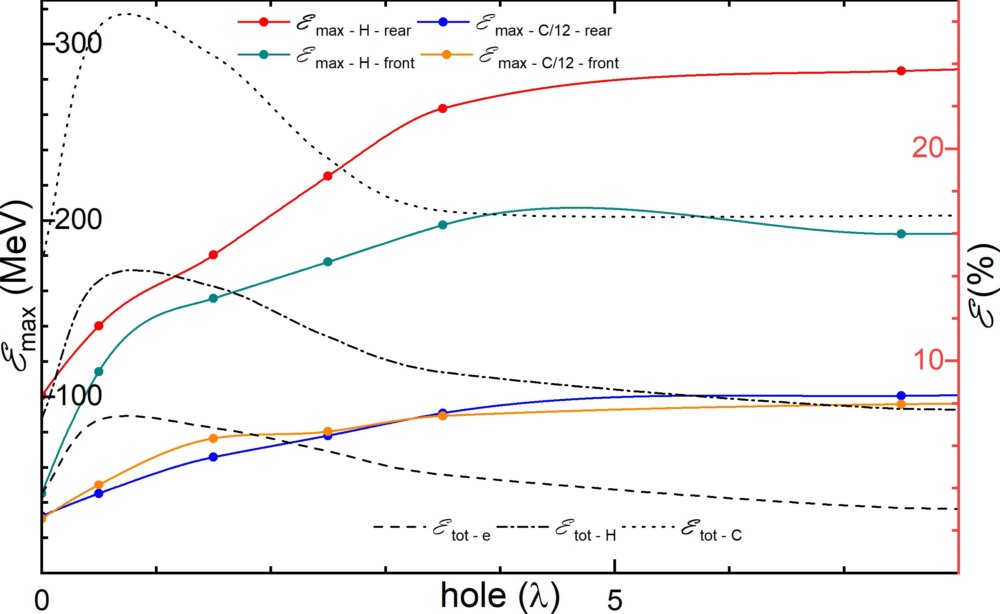}
  \caption{The target corresponds to a generalisation of the HT2SH. By increasing the side-holes extent and keeping the inner-segment constant, the target transits from a flat-foil to a ML target. (\textit{Left}) The maximum proton and carbon ion energy per nucleon for both the target rear and front, as indicated by the labels in the figure. (\textit{Right}) The percentage of the laser energy transferred to electrons, protons and carbon ions is indicated by the dash, dash-dot and dot line respectively.}
\label{fig:9_ML}
\end{figure}

\par Although the concept of ML targets is very attractive due to promising simulation results on enhancement of the maximum proton energy \citep{Limpouch2008, Bulanov2010a, Kluge2010, Lezhnin2016}, an experimental verification and implementation of the method is challenging \citep{Sokollik2010, PaaschColberg2011}. However, a ML foil is the limit of a HT2SH, when the extent of the side-holes tends to infinity. A summary of the simulation results is shown in Fig. \ref{fig:9_ML}, where the inner-segment has $1 \lambda$ extent.

\par At the other side of the limit, where the side-holes have a zero extent, the target corresponds to a flat-foil. There, the target rear protons have an energy of ${\sim} \kern0.1em 100 \kern0.2em \mathrm{MeV}$. However, the addition of two side-holes of ${\sim} \kern0.1em 2 \lambda$ doubles the initial energy, while expansion of the side-holes to infinity approximately triples the energy. Similar observations occur for the target front protons, where a flat-foil energy of ${\sim} \kern0.1em 50 \kern0.2em \mathrm{MeV}$ becomes four times higher when the side-holes tend to infinity. The same qualitative results also apply for carbon ions, where the initial ${\sim} \kern0.1em 35 \kern0.2em \mathrm{MeV}$ per nucleon are tripled, for both the target front and rear. In conclusion, the concept of a ML target gives approximately the same maximum proton energy with a HT2SH with sufficiently large side-holes, of ${\sim} \kern0.1em 5 \lambda$ extent.

\par As it is mentioned for Fig. \ref{fig:5a}, the holed-target geometry significantly increases the transfer of laser energy into particles. This effect can also be seen by the black lines in Fig. Fig. \ref{fig:9_ML}, where the dashed, dashed-dotted and dotted lines correspond to the percentage of the laser energy transferred to electrons, protons and carbon ions respectively; notably, the trend for all three particles is very similar regardless the value of the side-holes. For the case of a flat-foil the overall particles' energy is ${\sim} \kern0.1em 25 \kern0.2em \%$, while the addition of two side-holes of ${\sim} \kern0.1em 1 \lambda$ approximately doubles the absorbed energy. However, further increase of the side-holes dimensions lowers the laser energy transformation since a larger portion of the pulse does not interact with the target. By increasing the side-holes to ${\sim} \kern0.1em 3.5 \lambda$ (equivalent to two standard deviations of the electric field) then only a ${\sim} \kern0.1em 5 \kern0.2em \%$ of the pulse interacts with the outer-segments and its contribution can be ignored. At that point, the laser energy transformation is solely due to the inner-segment, saturating at an energy conversion efficiency of ${\sim} \kern0.1em 28 \kern0.2em \%$. Therefore, one must compromise between the highest proton energy obtained from ML targets and the maximum laser energy conversion efficiency obtained from HT2SH with holes of ${\sim} \kern0.1em 1 \lambda$.


\section{Summary and Conclusions}

\par Microstructured targets have gained significant popularity on laser-driven particle acceleration with many target designs already been proposed. In the present work an innovative target geometry is introduced, where three holes are created on a flat-foil. This general holed-target geometry has two limits, where at the first limit the inner-segments can be joined together, leaving only two side-holes into the target and is named as HT2SH. At the second limit the inner-segments are joined to the outer-segments, leaving only a central-hole on the foil, named as HT1CH. The length sum of the three holes plus the inner-segments is constant at  $8 \lambda$, corresponding to the electric field FWHM of the laser pulse. By plotting the contour of the side-holes and the central-hole extent, all geometrical combinations of the holed-targets are represented in an orthogonal triangular area, where the orthogonal angle of the triangle corresponds to a flat-foil which is the common geometry of the two limit groups.

\par The contour plots for the laser energy transformed into particle energy reveal that the holed-target geometry significantly enhances the laser energy conversion efficiency, where for some cases the improvement is approximately double compared to flat-foils. The different particles consisting the polyethylene target used, do not have optimum conversion efficiency for the same holed-target geometry; electrons have an optimum at a ${\sim} \kern0.1em 1 \lambda$ central-hole and ${\sim} \kern0.1em 0.5 \lambda$ side-holes, while protons at ${\sim} \kern0.1em 0.5 \lambda$ central-hole and ${\sim} \kern0.1em 2.5 \lambda$ side-holes.

\par The holed-targets take advantage of the laser pulse reflection, diffraction and interference to manipulate the electron energy spectra and their spatial distribution. If a hot electron population is captured by the laser pulse then it is continuously accelerated until it is dephased. In the case of a flat-foil, most of the pulse is reflected by the foil front surface and the captured electrons are accelerated in a direction opposite to the laser pulse. On the other hand, the holed-target geometry diffracts the pulse on the target rear region, where the electrons are now favourably accelerated. In addition, the electron temperature is significantly enhanced by holed-targets. However, this optimised behaviour ceases if the holes become extremely large, because the pulse edges do not sufficiently generate hot electrons at the initial stages of the interaction.

\par The maximum proton energy contour plots for the target rear create a pattern where two local maxima are observed. These two maxima correspond to the two limiting cases of the general holed-target geometry, where the highest is for the HT2SH. The dominant acceleration mechanism on that limit is the CE acceleration, where by reducing the dimensions of the inner-segment to an optimum value, protons of ${\sim} \kern0.1em 260 \kern0.2em \mathrm{MeV}$ are emitted. This value is higher than the ${\sim} \kern0.1em 180 \kern0.2em \mathrm{MeV}$ predicted for CE, indicating that the acceleration is slightly enhanced by a secondary mechanism. This secondary mechanism is the main acceleration mechanism for HT1SH. There, due to diffraction of the laser pulse, a longitudinal electric field is created which is responsible for the ${\sim} \kern0.1em 220 \kern0.2em \mathrm{MeV}$ proton energy observed. In contrast with CE, only low energy target front protons are emitted.

\par If the length sum of the side-holes and the inner-segments is not constant then another limiting case is considered by generalisation of the HT2SH (which is a sub-case itself), which is the transition from a flat-foil to a ML target. This transition occurs by keeping the central-foil constant and gradually increasing the extent of the side-holes. It is found that the energy absorption is optimised for side-holes of ${\sim} \kern0.1em 1 \lambda$ and it rapidly drops as the transition to ML targets occurs, due to less available particles in the laser focal spot. On the other hand, the maximum proton energy is significantly increased for increased side-holes extent, where a near-maximum value is obtained for side-holes of ${\sim} \kern0.1em 4 \lambda$, indicating that the ML targets have almost no benefit on particle acceleration compared to holed-targets. Therefore, a compromise must be made regarding optimisation of both conversion efficiency and maximum proton energy.

\hfill \break
\par The authors acknowledge fruitful discussions with P. Sasorov and T. M. Jeong, and thank A. Molodozhentsev for providing a significant part of the computational resources used. This work is supported by the project High Field Initiative (CZ.02.1.01/0.0/0.0/15\_003/0000449) from the European Regional Development Fund. It was also in part funded by the UK EPSRC grants EP/G054950/1, EP/G056803/1, EP/G055165/1 and EP/M022463/1.


\bibliographystyle{jpp}
\bibliography{jpp-bibliography}

\end{document}